# Interface Capturing Flow Boiling Simulations in a Compact Heat Exchanger

**Anna Iskhakova[1]**
North Carolina State University
Raleigh, NC, 27607
aiskhak2@ncsu.edu

**Yoshiyuki Kondo**
Mitsubishi Heavy Industries, Ltd.
Japan
yoshiyuki.kondo.th@mhi.com

**Koichi Tanimoto**
Mitsubishi Heavy Industries, Ltd.
Japan
koichi.tanimoto.wh@mhi.com

**Nam T. Dinh**
North Carolina State University
Raleigh, NC, 27607
ntdinh@ncsu.edu

**Igor A. Bolotnov**
North Carolina State University
Raleigh, NC, 27607
igor_bolotnov@ncsu.edu

**ABSTRACT**

*High-fidelity flow boiling simulations are conducted in a vertical minichannel with offset strip fins (OSF) using R113 as a working fluid. Finite-element code PHASTA coupled with level set method for interface capturing is employed to model multiple sequential bubble nucleation using transient three-dimensional approach. The code performance is validated against experiments for a single nucleation site in a vertical rectangular channel. To test the code performance, the studies for a bubble departing from the wall in a minichannel*

---

[1] Corresponding author

*with OSF are carried out first. Due to low heat flux values applied to the channel (1 kW/m²) contribution from the microlayer is not considered. The influence of surface characteristics such as contact angle and liquid superheat on bubble dynamics are analyzed. Local two-phase heat transfer coefficient is also investigated. To achieve higher void fractions, two conic nucleation cavities are introduced in the same channel with OSF. Observed bubble characteristics (departure diameter, bubble departure frequency) are evaluated and bubble trajectories are presented and analyzed. Local heat transfer coefficient is evaluated for each simulation case. The results show approximately 2.5 times increase in the local heat transfer coefficient when individual bubbles approach the wall. With a smaller bubble nucleation diameter, heat transfer coefficient increases by two times. The current work shows the capability of modeling flow boiling phenomena in such complex geometry as OSF as well as data processing advantages of high-resolution simulations.*

## 1. INTRODUCTION

Forced convection combined with boiling allows effectively remove heat from surfaces. High heat transfer coefficients explain a wide range of applications of flow boiling phenomenon: nuclear reactors, steam generators, microchannels, compact heat exchangers, etc. In the latter, fins of different shapes (rectangular, triangular, wavy, offset strip fin (OSF), etc.) might be added to increase heat transfer from the surface. In order to operate at low temperatures / pressures, refrigerants have been used as working fluids in compact heat exchanges [1, 2]. However, accurate prediction of two-phase heat transfer coefficient in such installations is still a challenge.

From experimental perspective, several studies have been conducted to predict two-phase heat transfer coefficient in minichannels with OSF. In [3] R113 behavior is examined in a vertical channel with OSF in a wide range of Reynolds numbers and applied heat fluxes. The authors proposed a correlation that allows to find two-phase forced convective part of local two-phase heat transfer coefficient. The results showed that measured in experiments heat transfer coefficient values lie within ±25% of the predictions. More recently an experimental study [4] was performed for R134a in a brazed heat exchanger with vapor quality up to 0.7. A correlation for Reynonds number factor $F$ was proposed (used for two-phase forced convective heat transfer coefficient evaluation) as well as two-phase frictional multiplier $\phi_f$ (applied in two-phase frictional pressure drop correlations). In [5] boiling heat transfer of HFE-7100 is investigated in a vertical channel with OSF and without them. The results showed that for low heat loads (less than 80 kW/m$^2$) convective boiling is a dominant process in two-phase heat transfer and, therefore, OSF play a major role. For higher heat loads (more than 120 kW/m$^2$) influence of the channel inner geometry is neglibigle, because nucleate boiling part starts to govern two-phase heat transfer. In [6] the authors conducted an experimental study with serrated fins in a horizontal rectangular channel using R134a as a working fluid. They varied heat flux and captured flow patterns from dispersed bubbly to annular flow followed by intermittent dry-out. The results showed that for a low vapor quality (less than 0.6) under high heat fluxes mass flux influence on heat transfer coefficient is almost neglibible, whereas for a high vapor quiality flow the heat transfer coefficient insreases with the mass flux growth.

Due to high computational cost, numerical investigations are mostly conducted to determine heat transfer coefficient for single-phase flows. In the work [7] the SST $k - \omega$ turbulence model is used to predict Colburn factor $j$ for a 3D geometry with OSF. The obtained results lie within ±20% range comparing to a well-known correlation proposed by Manglik and Bergles [8]. A year later the same researchers [9] proposed a new correlation for $j$ that includes Prandtl number. This was needed since the existing correlations were limited to a working fluid with a low $Pr$ (e.g., for air $Pr$=0.7). In [10] it was found out that in laminar flow heat transfer coefficient is mainly influenced by the growth of the boundary layer and in turbulent flow vortex shedding becomes important too. 3D simulations using Fluent was performed in [11] in which standard $k - \epsilon$ model used away from the walls and the SST $k - \omega$ turbulence model is applied near the walls. The authors established that a high difference in data (15%) appeared for $Re$ < 120 and around 35% difference in transition (500 < $Re$ < 1250) and turbulent ($Re$ > 1250) regimes compared to the expression from [8]. In both [12] and [13] 3D simulations in Fluent were conducted and compared with the existing correlation [8]. To cover new geometry parameters [12] / a wider range of $Re$ [13], new expressions for $j$ are proposed.

To model boiling flows with interface capturing, researchers employ various interface-tracking methods (ITM). For example, a volume of fluid (VOF) based flow solver was proposed in [14] that allows to reduce computational cost by keeping interfacial temperature at saturation and explicitly coupling it with source terms. This solver is validated, and its capability is demonstrating on 3D flow boiling simulation in a rectangular microchannel. In work done by [15] a new ITM is used that couples both VOF and level set methods (VOSET). Researchers implemented microlayer model to take into consideration the evaporation in the microlayer underneath a bubble and conjugate heat transfer between the wall and fluid region. They investigated subcooled boiling of water in a vertical rectangular channel in 2D. To ensure multiple bubble nucleation, 15 nucleation cavities are introduced in the domain. Varying heat fluxes from 100 to 500 kW/m$^2$ allowed them to observe flow regime transition from bubbly to elongated bubbly flow. In [16] the influence of surface characteristics on bubble dynamics is investigated in a vertical rectangular channel. Bubble departure from a solid surface is modeled in 2D. The authors concluded that hydrophilic surfaces allow to increase heat transfer from the wall due to evaporation of microlayer. On ambiphilic surfaces waiting time (period between the departure of one bubble and nucleation of a consequent bubble) is eliminated which results in effective contact line heat transfer. Numerical simulations conducted alongside with flow boiling experiments are presented in [17]. In this study FC-72 is used as working fluid in a vertical rectangular channel in 2D. The presented computational results showed good agreement with experimental data (including flow visualizations, cross-section averaged void fraction, velocity profiles, wall temperature and heat transfer coefficient). Although this prediction is worsened as mass velocity increases (there are model limitations to represent nucleate boiling in 2D). In [18] the authors investigated flow boiling of FC-72 in a vertical rectangular microchannel in 3D. The results revealed good agreement of numerical predictions for wall temperature variations and overall flow patterns with experimental results. This study also allowed to obtain fluid temperature distribution that cannot be measured in experiments with such small channel.

From the literature review it could be seen that due to high complexity of flow boiling phenomenon, high-resolution simulations are being done in 2D, with only a few cases conducted in 3D. Additionally, studies presented here consider a simple rectangular geometry of a channel. To provide a deeper understanding of phenomena happening in flow boiling in a vertical minichannel with OSF, the current studies are conducted. Finite-element code PHASTA is employed to solve Navier-Stokes equations (section 2). The code performance is validated against experiments for a single nucleation site in a vertical rectangular channel [19, 20] (section 3). Then the studies for a bubble departing from the wall in a minichannel with OSF are carried out (section 4). Influence of surface characteristics such as contact angle and liquid superheat on bubble dynamics are analyzed. The results show that there is a small difference in bubble departure diameters (~4%) for contact angles 30° and 45°. Local two-phase heat transfer coefficient is also investigated. To achieve higher void fractions, two nucleation cavities are introduced in the same channel with OSF (section 5). For these studies liquid superheat influence on bubble dynamics is also evaluated. Heat transfer enhancement due to bubble presence is quantified on planes along the bubbles' path.

## 2. NUMERICAL TOOLS

### 2.1. PHASTA Description

Finite element flow solver PHASTA (**p**arallel, **h**ierarchic, higher-order accurate, **a**daptive, **s**tabilized, **t**ransient **a**nalysis) is used for the current simulations. The code solves incompressible Navier-Stokes equations in the strong form that are

$$\frac{\partial u_i}{\partial x_i} = 0 \qquad (1)$$

$$\rho \frac{\partial u_i}{\partial t} + \rho u_j \frac{\partial u_i}{\partial x_j} = -\frac{\partial p}{\partial x_i} + \frac{\partial \tau_{ij}}{\partial x_j} + f_i \qquad (2)$$

$$\rho c_p \left( \frac{\partial T}{\partial t} + (\vec{u} \cdot \nabla) T \right) = \nabla \cdot (k \nabla T) + \vec{q} \qquad (3)$$

where $u_i$ is the fluid velocity, $\rho$ is the density, $p$ is the static pressure, $\tau_{ij} = \mu \left( \frac{\partial u_i}{\partial x_j} + \frac{\partial u_j}{\partial x_i} \right)$ is the viscous stress tensor, $f_i$ is the body force, $c_p$ is the specific heat at the constant pressure, $T$ is the temperature, $k$ is the thermal conductivity of a fluid and $\vec{q}$ is the dissipation function. For surface tension, the Continuum Surface Force (CSF) model is used introduced by [21]. Discretization schemes that are implemented in PHASTA are described in more details in [22, 23].

To capture the interface, the level set method is employed [24]. The method introduces a distance function $\varphi$ that defines the shortest distance to the interface from any point in the domain. At the interface $\varphi = 0$, in the vapor phase $\varphi < 0$ and in the liquid phase $\varphi > 0$. The distance function is advected by the flow velocity as

$$\frac{\partial \varphi}{\partial t} + u_i \frac{\partial \varphi}{\partial x_i} = 0 \qquad (4)$$

To decrease a jump in properties across the interface, a smoothed Heaviside function $H_\varepsilon$ is applied in which interface thickness $\varepsilon$ is used as

$$H_\varepsilon(\varphi) = \begin{cases} 0, & \varphi < -\varepsilon \\ \frac{1}{2} \left[ 1 + \frac{\varphi}{\varepsilon} + \frac{1}{\pi} sin\left( \frac{\pi \varphi}{\varepsilon} \right) \right], & |\varphi| < \varepsilon \\ 1, & \varphi > \varepsilon \end{cases} \qquad (5)$$

The Navier-Stokes equations are solved using 'one-fluid' approach for which fluid properties (density, viscosity, specific heat and thermal conductivity) are found using $H_\varepsilon$ (e.g., $\rho(\varphi) = \rho_l H_\varepsilon(\varphi) + \rho_g (1 - H_\varepsilon(\varphi))$, where $\rho_l$ and $\rho_g$ are the liquid and vapor phase densities respectively).

## 2.2. Contact Angle Control Algorithm

Contact angle control algorithm [25] is employed in the current simulations to represent surface characteristics. For this algorithm, a target contact angle is needed to be specified based on experimental data. Spatially varying subgrid force $F_{CA}$ is applied at the wall-adjacent cells to ensure that the current contact angle is equal to (or in a range of) the desired value.

$$F_{CA} = F_1 cos \left( \frac{\varphi}{T_{CA}} \cdot \frac{\pi}{2} \right) \rho (H_{CA} - d_{wall})^2 \tag{6}$$

In the above expression $F_1$ is the parameter that changes $F_{CA}$ with respect to the desired contact angle value [26], $T_{CA}$ and $H_{CA}$ are the thickness and height of $F_{CA}$ application region respectively, $d_{wall}$ is the distance from the wall. Current contact angle is calculated using interface normal vector (which is found using distance function as $\nabla \varphi$) and wall normal vector in wall-adjacent elements. More details on contact angle algorithm implementation in PHASTA and its verification could be found in [25].

## 2.3. Evaporation and Condensation Algorithm

In the presented flow boiling simulations Evaporation and Condensation algorithm is used that was previously introduced in PHASTA [27]. In this algorithm it is assumed that the vapor phase temperature inside a bubble is constant (and equals the saturation temperature for the current conditions). Average temperature gradient is found for each bubble using Bubble Tracking Algorithm [28]. Local heat flux across the interface is calculated as

$$q'' = -\left(\frac{R_0}{R_1}\right)^2 \cdot k_l \frac{\partial T}{\partial n} \qquad (7)$$

In the above equation $R_0$ is the bubble radius, $R_1$ is the radius of temperature gradient collection shell [27], $k_l$ is the liquid phase thermal conductivity, $\frac{\partial T}{\partial n}$ is averaged temperature gradient across the interface. Volume change that is later applied to the RHS of the continuity equation (1) is found as

$$dV = \frac{4\pi R_0 \bar{q}}{h_{fg}(\rho_l - \rho_g)} \qquad (8)$$

Here $\bar{q}$ is the volume-averaged heat flux (this procedure is needed to ensure the correct code behavior on unstructured meshes), $h_{fg}$ is the latent heat of evaporation. This algorithm was verified against analytical bubble growth model [29] and experimental data for pool and flow boiling [30].

## 3. MODEL VALIDATION

In order to validate the performance of Boiling and Contact angle algorithms for flow boiling conditions, a separate study is conducted. Experiments in a vertical rectangular channel with HFE-301 are chosen as the reference [19, 20]. These experiments were conducted for the single nucleation site and such parameters as bubble departure diameters and departure frequencies are available for comparison.

The experiments [19, 20] were carried out in a wide range of parameters. However, to match the desired flow conditions (will be described in next sections), the following flow characteristics are considered: mass flux 140 kg/m2·s ($Re$ = 3125), wall heat flux 9.7 kW/m$^2$, inlet subcooling 13.5 K, atmospheric pressure.

Simulation domain constructed for this study is shown in Fig. 1. The channel dimensions are 324x10.1x10 mm. It has a total of 264 mm before the test section (including 50 mm unheated and 214 mm heated parts of the channel). The test section is 10 mm long and has a nucleation site of a trimmed conic shape placed in the middle of the section. Additional 50 mm are added after the test section for outlet treatment (to avoid backflow issue and dissipate bubbles since mesh is comparatively coarse there). The heated area in simulations matches the experimental one.

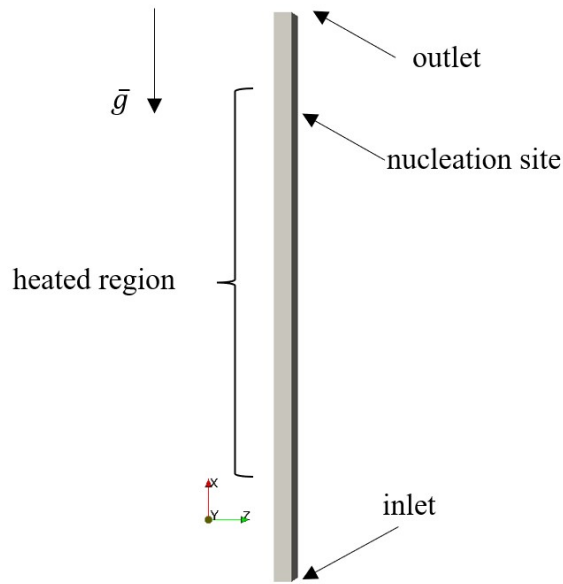

Fig. 1. Computational domain for validation studies

The real cavity size is in order of nm which is very computationally expensive to model. Instead, three bigger cavity sizes are considered: 0.288, 0.18 and 0.1125 mm of the cavity diameter (the depth is the same for all three cases 0.128 mm) (see Fig. 2). The mesh used near cavities will be recycled for OSF simulations (discussed in next sections).

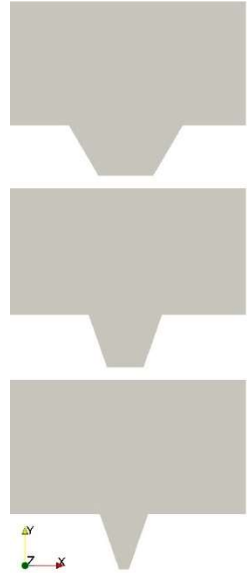

Fig. 2. Different cavity sizes considered: baseline (top), 38% smaller (middle) and 61%

smaller (bottom)

Exact surface characteristics are unknown [19, 20]. However, using information about measured contact angle for the refrigerant with similar properties [31] as well as magnified views of high-resolution camera [20], contact angle 15º is set for these validation studies.

Bubble departure diameter and departure frequency are compared for three

validation cases in Fig. 3, top. The waiting time is part of the fully resolved simulation and

not specified by any modeling in our results. However, limited number of nucleation sites

results in large superheated area around the nucleation sites, and thus the resolved

waiting time is fairly small. Departure frequency is calculated as

$$f_i = \frac{1}{t_{d_i} - t_{n_i}}, \tag{9}$$

where $f_i$ bubble departure frequency for $i^{th}$ bubble, Hz; $t_{d_i}$ time of departure of

$i^{th}$ bubble, s; $t_{n_i}$ time of nucleation of $i^{th}$ bubble, s.

It could be seen that as the cavity diameter is decreasing, the departure diameters of bubbles become smaller. This happens since the cavity is always filled with vapor and thus, bubble interface is attached to the cavity's edges (constant boundary conditions are set for level set function on cavity walls). These boundary conditions are needed to ensure multiple bubble nucleation. Otherwise, there would be no vapor phase left in the cavity. With this treatment, with a smaller cavity diameter a bubble is prone to depart sooner.

The results for the smallest cavity agree well with experimental data [20]. From bubble departure frequency plot (Fig. 3, top right) one may see that the data is consistent: smaller bubbles depart faster from the cavity. It could be noticed that the modeled values are somewhat lower than that observed in experiments (e.g., $MSE$ for the smallest cavity is 146.2 Hz).

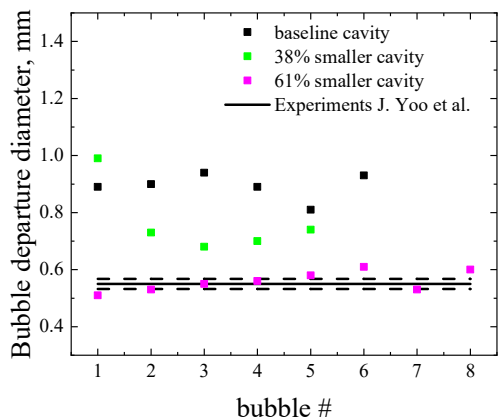
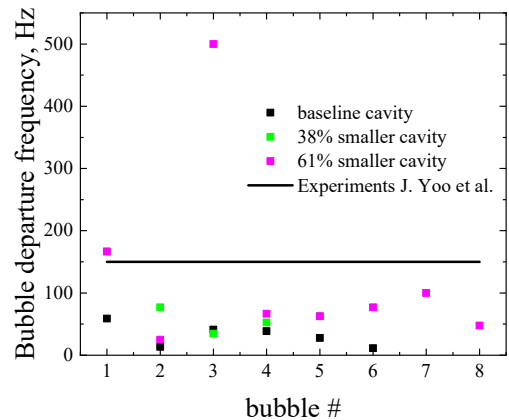

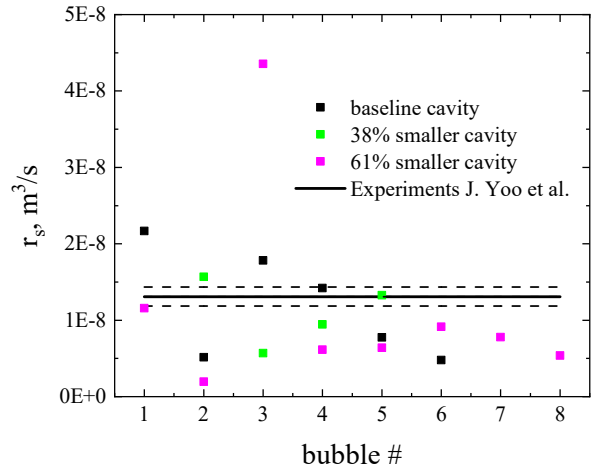

Fig. 3. Bubble statistics are compared with experiments [20] (dash lines on the left plot

define uncertainty in departure diameter measurements)

To better understand this, the amount of steam released from a cavity could be found as

$$r_s = \frac{V}{t_b},$$ (10)

where $V$ is bubble volume ($\frac{\pi D^3}{6}$), m³; $t_b$ is the nucleation period (time difference between the departure and the nucleation for a bubble), s. The resulted values are shown in Fig. 3, bottom. From this plot it could be seen that the amount of steam is consistently lower for the smallest cavity (as bubble departure frequency results discussed before). This could be attributed to microlayer evaporation that is not considered in this study. In the future work additional source term could be considered to account for the amount of steam generated in this tiny liquid film close to the wall.

## 4. BUBBLE DEPARTURE FROM THE WALL

### 4.1. Model Setup

To model the flow through OSF, a part of a vertical rectangular channel used in experiments [3] is considered (Fig. 4, left). In the domain only flow region is modeled, therefore there is a 'step' of 0.2 mm corresponding to the fin thickness (Fig. 4, right) that is present in $y$ direction. The flow is resolved for the first 12 mm of the domain, other parts are created to increase the code robustness (coarse meshes are used to prevent backflow and dissipate bubbles, the outlet treatment box includes last 6 mm (not shown in Fig. 4, right)).

No-slip boundary conditions are imposed at the top and bottom of the domain, the flow is periodic in $z$ direction. Inflow / outflow boundary conditions are set for velocity and temperature profiles in $x$ direction. Gravity is applied in the opposite direction of the flow. There is an uncertainty in the velocity and temperature profiles in the channel. To deal with it, it is assumed that the flow is fully developed in the modeled part of the channel. Also based on experimental data [3], $Re$ = 200 is chosen for these simulations. Therefore, parabolic velocity and temperature profiles are considered to be representative for this flow. Constant temperature $T_w$ = 323.3 K (2 degrees superheat) is set at the right and left walls of the domain (which represent inner channel walls assuming highly conductive fin material). According to FLUENT simulations [32], this value corresponds to heat flux 1000 W/m$^2$ that was used in the experiments [3]. Since there is an uncertainty in exact temperature values at the inner walls, superheat studies are performed. Fluid properties for R113 at 102.4 kPa are presented in Table 1. Initial time step is set 5·10$^{-6}$ s and was automatically adjusted during the simulation to keep Courant–Friedrichs–Lewy ($CFL$) number = 1.0.

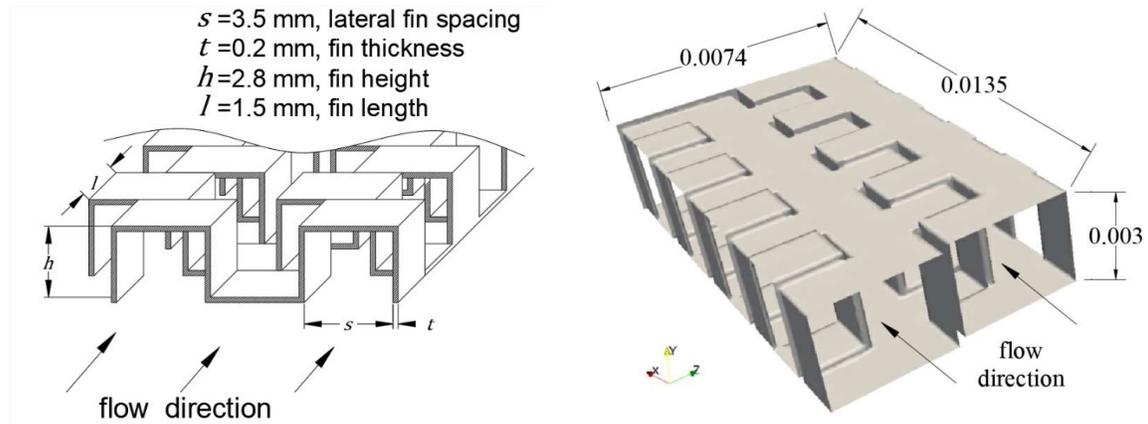

Fig. 4. Geometry of OSF (left), corresponding computational domain (right, dimensions are shown in m)

Table 1. Fluid properties

| Name | Liquid phase values | Vapor phase values |
|---|---|---|
| Dynamic viscosity, Pa·s | $5 \cdot 10^{-4}$ | $1.1 \cdot 10^{-5}$ |
| Density, kg/m³ | 1507 | 7.5 |
| Specific heat, J/(kg·K) | 941 | 692 |
| Thermal conductivity, W/(m·K) | $67.4 \cdot 10^{-3}$ | $9.51 \cdot 10^{-3}$ |
| Surface tension coefficient, (N/m) | 0.01456 | |

## 4.2. Mesh Design

In such type of a complex geometry, considerable friction losses are present when the flow goes through OSF. In order to properly resolve these losses, boundary layer meshing structure (BLs) is created on wall surfaces (Fig. 5). To accurately resolve a bubble (around 25-30 elements across the bubble diameter are needed [33]) as well as to minimize the computational cost, several mesh refinement regions are introduced (Table 2). The bubble with initial radius 0.3 mm is placed at the location (0.00825; 0.000135; 0.004775). This size is chosen to save computational resources (a smaller bubble would require a higher mesh resolution). The bubble is located at the bottom step of the fin. Due to stagnant liquid at the step corners, more superheated liquid is accumulated here.

Therefore, bubble departure diameter is expected to be bigger rather than when the bubble is initialized at the top of the fin. Target contact angle 30° is set. This value is a source of uncertainty (since exact surface characteristics are unknown), therefore, contact angle studies are performed. For the contact angle algorithm to perform as designed, BLs are needed on the wall (Fig. 5, bottom).

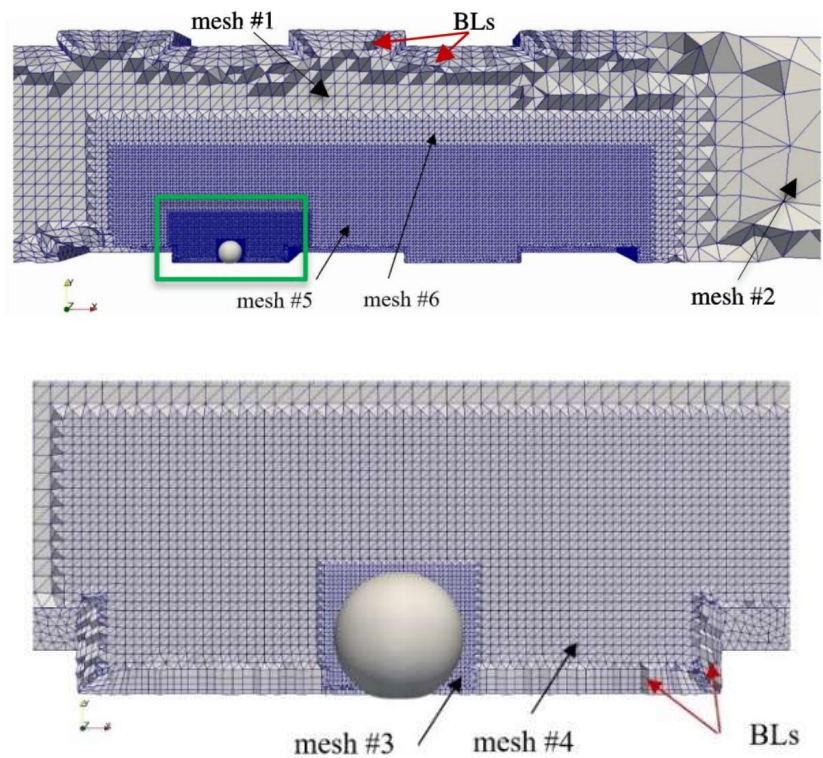

Fig. 5. Mesh refinement regions (top), magnified view of the mesh near the initialized bubble (bottom)

Table 2. Mesh sizes and number of elements for each mesh

| # | Mesh resolution, m | Presence of BLs | Number of elements |
|---|---|---|---|
| 1 | $1.6 \cdot 10^{-4}$ | yes | 707,092 |
| 2 | $1.28 \cdot 10^{-3}$ | no |  |
| 3 | $1.0 \cdot 10^{-5}$ | no | 410,502 |
| 4 | $2.0 \cdot 10^{-5}$ | yes | 793,797 |
| 5 | $4.0 \cdot 10^{-5}$ | no | 1,984,200 |

| 6 | 8.0·10⁻⁵ | no | 472,760 |
|---|---|---|---|

### 4.3. Initial Conditions for Boiling Simulations

It was mentioned before that parabolic velocity and temperature profiles are imposed at the channel inlet to achieve laminar ($Re$ = 200) flow. However, due to the staggered geometry of the channel (see Fig. 5, top), initially these profiles are somewhat different near the right and left walls. To overcome this discrepancy, single-phase simulations are conducted first to make sure that temperature and velocity gradients are consistent near the walls. Simulations are conducted for 1.7 s with maximum $CFL$ number allowed 20.0. The resulted velocity and temperature profiles are shown in Fig. 6.

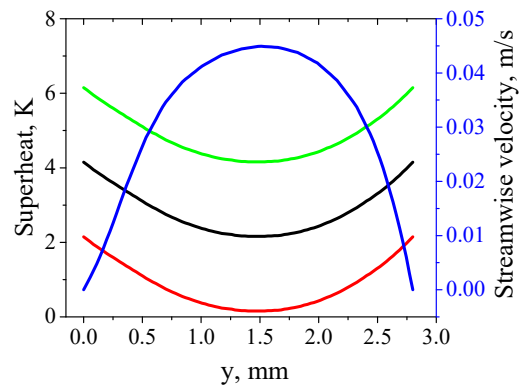

Fig. 6. Velocity (blue line) and temperature (green, black and red lines) profiles applied for three superheat cases

Single-phase simulations are carried out for all three bulk superheats considered (0, 2 and 4 degrees). Then the solutions at $t$ = 1.7 s are set as initial conditions (ICs) for flow boiling simulations.

### 4.4. Contact Angle Parametric Studies

Contact angle is an important parameter for boiling simulations. To investigate its influence on bubble growth and departure, two contact angles are considered: 30° and 45°. Bubbles at initial time step are presented in Fig. 7, top. The resulted bubble characteristics are shown in Table 3. Using Bubble Tracking Algorithm [28], bubble growth statistics are obtained (see Figure S2, available in Supplemental Material, part of the ASME Digital Collection).

Table 3. Contact angle studies parameters

| Case # | Contact angle value | Superheat at the wall (in the channel center), K | Bubble departure diameter, m | Departure time, ms |
|--------|--------------------|--------------------------------------------------|------------------------------|---------------------|
| 1 | 30° | 2 (0) | $4.113 \cdot 10^{-4}$ | 5.16 |
| 2 | 45° | 2 (0) | $3.964 \cdot 10^{-4}$ | 4.37 |

Contact angle 30°                    Contact angle 45°

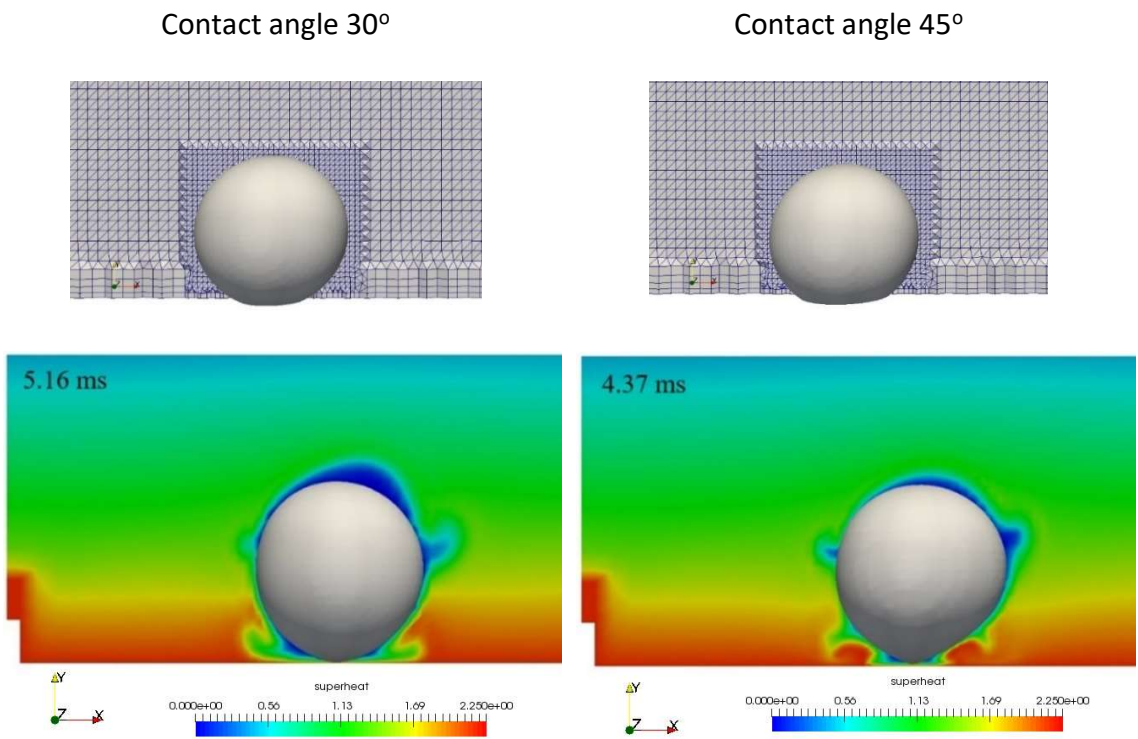

Fig. 7. Bubbles at initial time step (top) and at the departure moments (bottom)

From Fig. 7 one may see that bubble with a higher contact angle reaches smaller departure diameter and departs faster. Overall, the difference in bubble departure diameters for two contact angles considered is ~4%. The results are obtained for the single bubble departure event only. To verify this observation, multiple bubble nucleation events are needed (to have more reliable bubble statistics). For that purpose, section 5 discusses flow boiling simulations conducted in the same geometry with OSF, but with 2 nucleation sites.

## 4.5.    Superheat Studies

Parabolic temperature profile as well as wall temperature are the sources of uncertainty in the current study. To investigate the influence of wall temperature values on bubble growth, three superheat studies are conducted (Table 4). The simulations were conducted until a bubble reaches the coarser mesh region (mesh #6, Fig. 5). The average distance from the nucleation site that the bubble covers is around 0.00375 m. Wall superheat is chosen in a way to keep the same temperature difference between the wall and the channel center. Mesh design and other model conditions are the same as in Contact angle studies. For the current simulations target contact angle is equal 30°, advancing contact angle is 25° and receding contact angle is 35°. The resulted bubble growth statistics are shown in Figure S3, bubble positions at the departure moments are presented in Figure S4, both are available in Supplemental Material, part of the ASME Digital Collection.

Table 4. Superheat studies parameters

| Case # | Contact angle value | Superheat at the wall (in the channel center), K | Bubble departure diameter, m | Departure time, ms |
|--------|---------------------|--------------------------------------------------|------------------------------|--------------------|
| 3 | | 2 (0) | $5.237 \cdot 10^{-4}$ | 10.62 |
| 4 | $30^{\circ}$ | 4 (2) | $8.126 \cdot 10^{-4}$ | 17.28 |
| 5 | | 6 (4) | $1.410 \cdot 10^{-3}$ | 27.09 |

Higher superheat values at the surface lead to a more intensive evaporation process. Because of that, bubbles reach bigger sizes for cases #4-5 compared to case #3. However, since vapor prevents liquid from coming closer to the surface, it worsens heat transfer for case #4-5 relative to case #3. On the other hand, the small-size bubble (as in case #3) departs much sooner from the wall and, thereby, heat transfer coefficient is increased. Another observation is that a larger bubble (cases #4-5) slides longer along the wall (see bubble departure locations in Figure S4, available in Supplemental Material, part of the ASME Digital Collection: in case #4 the bubble slides to the fin corner and then departs; in case #5 the bubble departs only in the middle of the next fin since it was sliding along the surface before that). Such bubble behavior tends to disturb thermal boundary layer, hence, increasing heat transfer coefficient [34, 35]. More details on heat transfer coefficient evaluation will be presented in the next sections.

**4.6.    Local Void Fraction Calculation**

Since only one bubble is present in each simulation, a smaller part of the channel is considered (Fig. 8, left) for local void fraction evaluation. The resulted void fraction statistics for each case are shown in Fig. 8, right. It could be noticed that since for case #5 the bubble size is larger, its void fraction correspondingly has higher values compared to cases #4 and 3. The bubble growth is somewhat similar for each case: as the bubble attaches to the surface, its volume is growing non-linearly. However, when the bubble departs and starts to move toward the channel center, the growth becomes more linear. Since for case #3 there is no superheat at the channel center, bubble growth rate in this region is considerably smaller than, for example, in case #4. For case #5 there is 4 K superheat at the channel center which, consequently, leads to a much higher bubble growth rate compared to cases #3 and 4. Final local void fraction achieved for cases #3-5 are 0.57%, 1.42% and 4.78% respectively.

Bulk superheat studies show that temperature distribution in the channel plays a crucial role in heat transfer evaluation. The difference in 2 degrees (cases #3 and 4) results in ~50% increase of the bubble departure diameter and the change in 4 degrees (cases #3 and 5) – in nearly 170% growth.

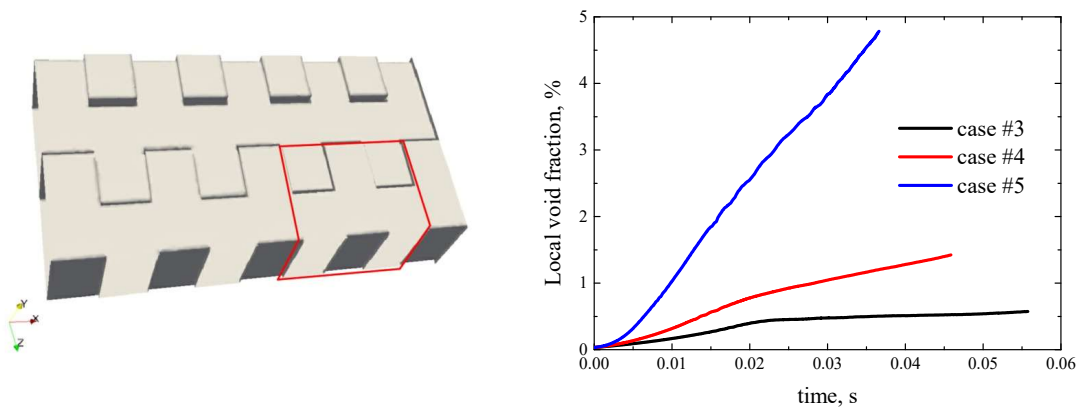

Fig. 8. Local void fraction evaluation region (in red) (left) and void fraction statistics for cases #3-5 (right)

### 4.7. Local Heat Transfer Evaluation

Local two-phase heat transfer coefficient is found for the cases #3-5 as

$$h = \frac{q''}{\Delta T} = \frac{-k}{\Delta T} \cdot \frac{1}{S} \int_S \nabla T dS. \tag{11}$$

In the equation above values are found from Fourier's Law consideration using area-averaged normal to the wall temperature gradient values ($\frac{1}{S}\int_S \nabla T dS$ in (11)), thermal conductivity of R113 liquid phase $k$ (see Table 1 for reference) and known temperature differences at the walls and the centerline ($\Delta T$ = 2 degrees). More details on that could be found in [36]. These temperature gradients are obtained at the 0.04 mm from the wall. This value is chosen to make sure that temperature changes linearly over this distance. The calculations are performed for each plane on the bubble path (see Figure S6, available in Supplemental Material, part of the ASME Digital Collection) at 4 time instances: 0, 9, 18 and 27 ms. Points at $t$ =0 s corresponds to single-phase heat transfer coefficients.

Due to less dense vapor phase, bubbles move faster than the liquid. Correspondingly, a bubble of a bigger size travels further downstream (see Figure S1, available in Supplemental Material, part of the ASME Digital Collection). The results show that local heat transfer coefficient for case #5 is generally higher than for cases #3-4. This could imply that nucleate boiling part of the two-phase heat transfer coefficient that depends on the liquid superheat plays an important role (for case #5 the superheat rate at the wall is the highest). At the same time, convective boiling heat transfer part is also significant for case #5: from snapshots of bubbles' location, one could notice a larger disturbance of the thermal boundary layer caused by the biggest bubble sliding along the wall (compared to cases #3-4).

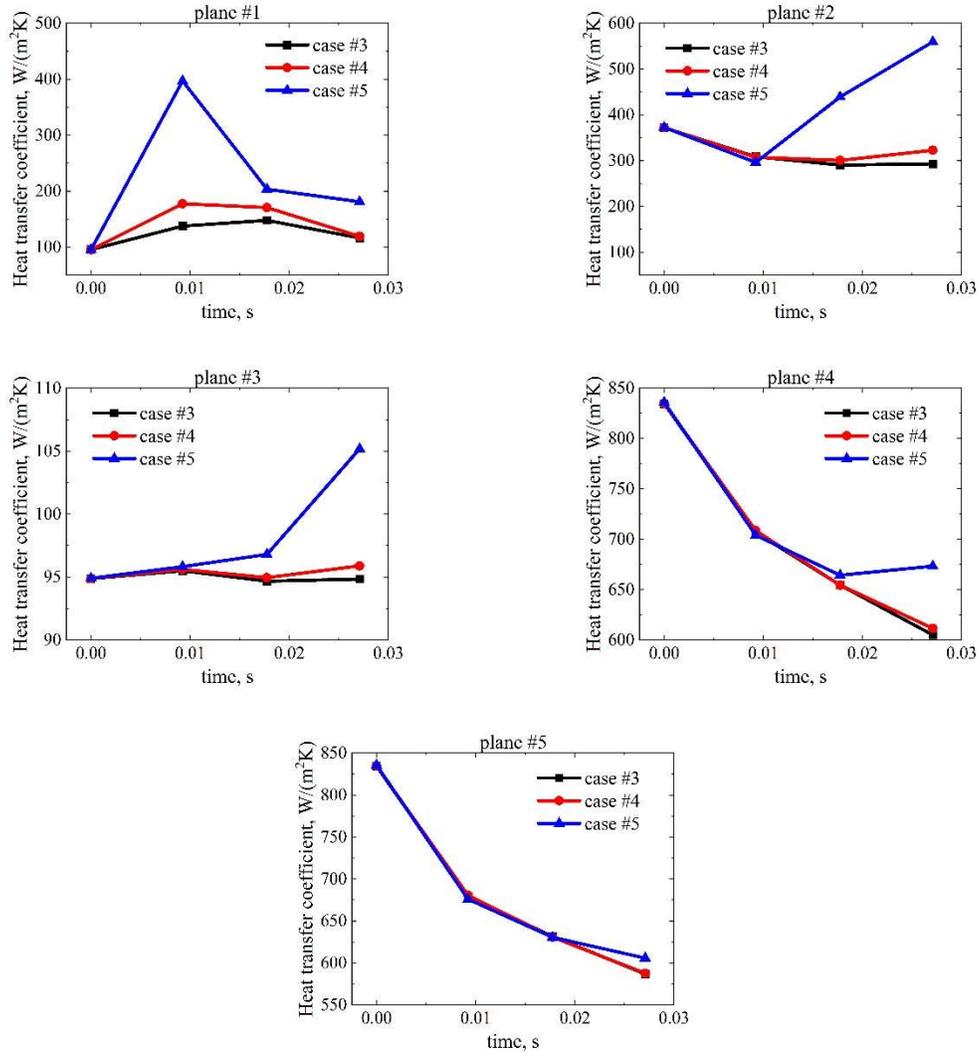

Fig. 9. Local heat transfer coefficient changes for different planes on the bubble path

By analyzing heat transfer coefficient change on plane #1 one could conclude that as the bubble grows, it acquires more energy due to phase change. Then as the bubble starts to move downstream following the flow velocity, it leaves the plane, and the heat transfer coefficient is correspondingly decreasing for all cases. When the bubble comes closer to plane #2, the heat transfer coefficient is increased significantly. It should be noted that since for cases #3-4 bubbles traveled smaller distance over this plane, the growth of the heat transfer coefficient is not as significant as for case #5. For plane #3 the heat transfer coefficient starts to increase as the bubble would closer approach this surface. For planes #4 and 5 (side walls) the heat transfer coefficient trend could be explained as follows: when the bubble is relatively far away from the side surface (as, for example, in cases #3 and 4), due to phase change it consumes the amount of liquid in the vicinity of the wall, thereby, degrading convective heat transfer. However, when the bubble comes closer to the wall (or its size is increasing), boiling heat transfer prevails. Observed local heat transfer coefficient values are obtained for the section of the channel where the bubble is nucleated. Overall, the heat transfer enhancement effect is more pronounced for plane #1 where the bubble is nucleated. Once it starts to grow (at $t =$9 ms), heat transfer coefficient is increased by 44.2% (case #3), 86.1% (case #4) and 315.7% (case #5).

In order to validate the results, the authors intend to compare it with the experiments [3] as the part of the future work. However, for this a higher void fraction should be achieved which imposes additional numerical challenges.

### 5. BUBBLE DEPARTURE FROM NUCLEATION CAVITIES

### 5.1. Model Setup

To increase void fraction, two nucleation cavities are introduced in the same channel on the right and left walls (see Figure S7, available in Supplemental Material, part of the ASME Digital Collection). Cavities have conic shape (diameter 0.288 mm, depth 0.128 mm) and are trimmed at the bottom. The two cavities are located at (8.25; 0; 4.675) and (9.75; 2.8; 4.675) mm. Exact locations of nucleation sites in experiments [3] are unknown as they depend on surface imperfections and may change for different heat flux conditions. Information on mesh used for these studies is available in Supplemental Material, part of the ASME Digital Collection.

As for the previous runs, since the exact temperature values at the inner channel walls likely depend on the streamwise location in the full heat exchanger design (in experiments the size of the channel is 750x3x100 mm), several simulations are conducted with three different bulk superheat values: 0, 2 and 4 degrees. These simulations are conducted with $Re$ = 200. For these simulations target contact angle is set 15°, advancing contact angle is 10° and receding contact angle is 20°. The same working fluid is considered R113 (Table 1). Simulations are carried out with the initial time step 2.1 microseconds, and it is adjusted during the runs to keep flow $CFL$ number = 0.5.

It has been observed that when a bubble grows at the wall, there is a thin liquid film exists between the bubble and the wall called microlayer [37, 38]. The contribution of the microlayer to bubble evaporation is substantial (heat fluxes reach 1 MW/m$^2$ [39]) especially at atmospheric pressure. However, as was shown in [40], heat flux from liquid evaporation in the microlayer region is negligible for low heat fluxes. Since in the current simulations applied wall temperature corresponds to heat flux 1000 W/m$^2$, the contribution from the microlayer is not considered.

Wall superheat plays a major role in the bubble growth, and it is important to accurately represent it in the simulations. There are several examples in which conjugate heat transfer between the solid and fluid domains was accounted for in boiling simulations, including [41, 42]. However, since in the current studies wall thickness is small (0.2 mm) and the material is highly conductive (copper) [3], conjugate heat transfer is neglected.

## 5.2. Superheat Studies

### 5.3.1. *Difference in Temperature Distribution*

To quantify the difference in the temperature distribution in the channel, sensitivity studies with three different bulk superheats are conducted. Departure diameters of the first-generation bubbles (e.g., the first bubble produced at each site) are summarized in Table 5. The case with 0 degrees superheat is considered as a baseline for this study. From Table 5 it could be seen that 2 degrees variation in bulk superheat causes a significant difference between departure diameters. This effect is even more pronounced for 4 degrees superheat difference. In Figure S5, available in Supplemental Material, part of the ASME Digital Collection, bubble shapes for the first-generation bubbles at the right wall cavity are shown at simulation time 6 ms. It could be seen that for 2 and 4 degrees bulk superheat bubbles are ready to depart, whereas for the 0 degrees bulk superheat case the bubble is more closely attached to the wall. Comparing to the bubble departure from the wall simulations (the results are shown in Table 4), current bubble departure diameters are smaller since a smaller contact angle (i.e., 15º) is imposed.

Table 5. Departure diameter of the first-generation bubbles

| Bulk superheat, K | Departure diameter, mm | | Average difference, % |
|---|---|---|---|
| | Nucleation site #1 | Nucleation site #2 | |
| 0 | 0.4535 | 0.453 | - |
| 2 | 0.5126 | 0.5122 | 13.05 |
| 4 | 0.6379 | 0.6329 | 40.19 |

### 5.3.2. Bubble Statistics

Some of the bubble statistics are shown in Fig. 10, bubble trajectories are available in Supplemental Material, part of the ASME Digital Collection. From these plots it could be seen that departure diameter is generally higher for the higher bulk superheat. For 2 and 0 degrees bulk superheat cases obtained departure diameter values are more consistent whereas for 4 degrees a higher discrepancy is observed (which is attributed to coalescence between just departed bubbles and bubbles left in cavities). Bubble departure frequency is more consistent when there is no coalescence between already departed bubbles and newly nucleated ones.

Snapshots of simulations for 4 degrees bulk superheat are shown in Fig. 11, for 0 and 2 degrees bulk superheats in Fig. 12. One could notice that for higher temperature gradients values near the walls, bubbles grow larger (for example, compare all three cases at 8 ms). For 4 degrees bulk superheat bubbles reach around 1 mm in diameter (more detailed data is presented below) and they extensively coalesce with each other (see snapshot for 32 ms, Fig. 11). At the end of this simulation vapor phase occupies more than a half of the channel cross section which could later results in a flow transition (from bubbly to plug flow).

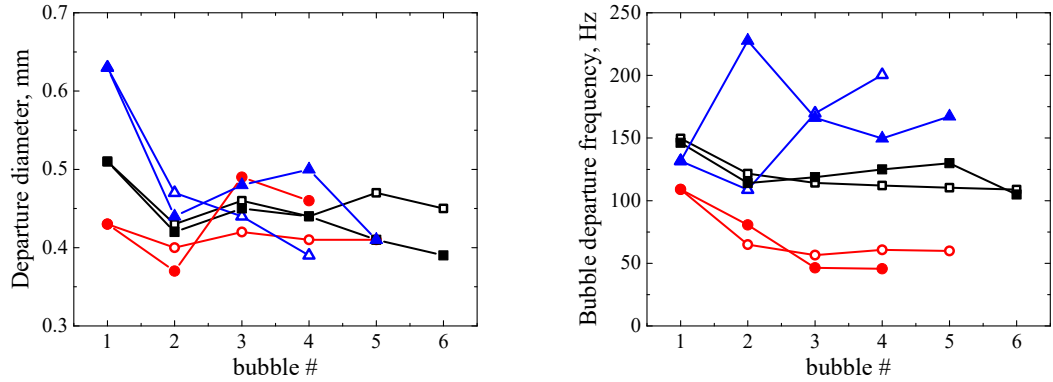

Fig. 10. Bubble statistics for nucleation site #1 (on the right wall, open symbols) and #2 (on the left wall, solid symbols) for 0 (red), 2 (black) and 4 (blue) degrees bulk superheat

Results shown in Fig. 12 indicate that for lower bulk superheat values (e.g., 0 degrees) bubbles have smaller departure diameters. Due to the size, bubbles coalesce less often and nearly stop to grow once they are near the centerline (there is no superheat here). For 2 degrees bulk superheat more bubbles are observed compared to the previous case due to higher temperature values in the channel.

4 degrees bulk superheat

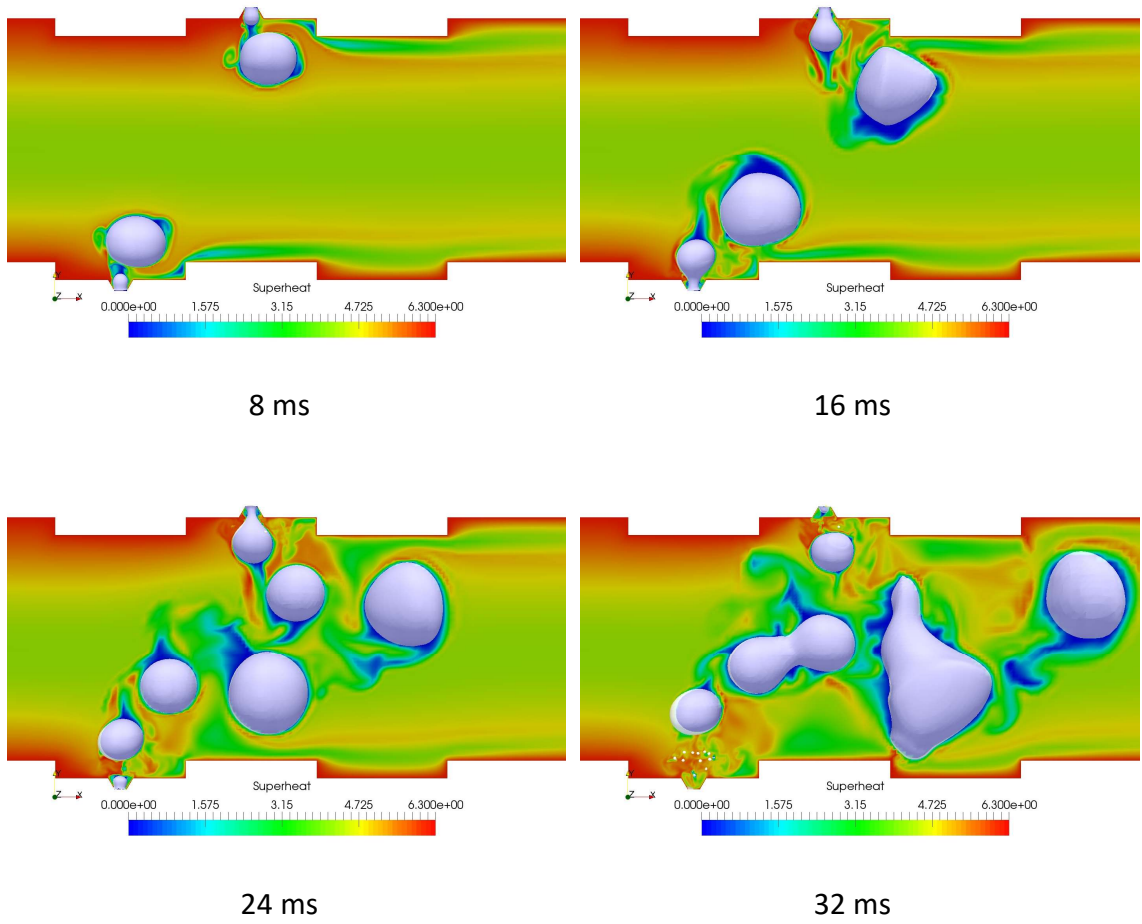

Fig. 11. Bubble dynamics for 4 degrees bulk superheat

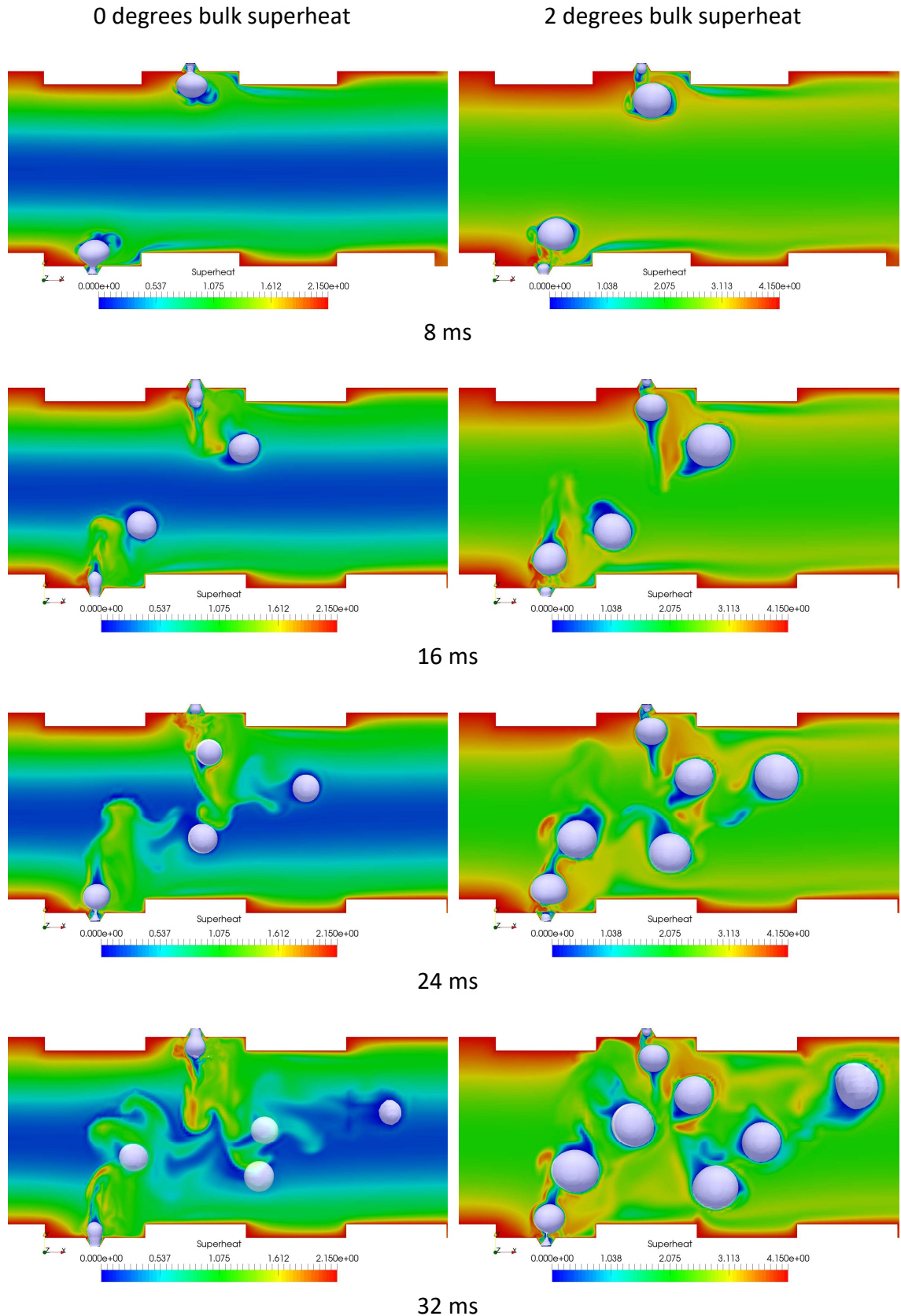

Fig. 12. Bubble dynamics for 0 and 2 degrees bulk superheat

### 5.3.3.  Local Heat Transfer Evaluation

Heat transfer coefficient is calculated locally on planes along bubbles' path (see Fig. 13) for 3 considered cases.

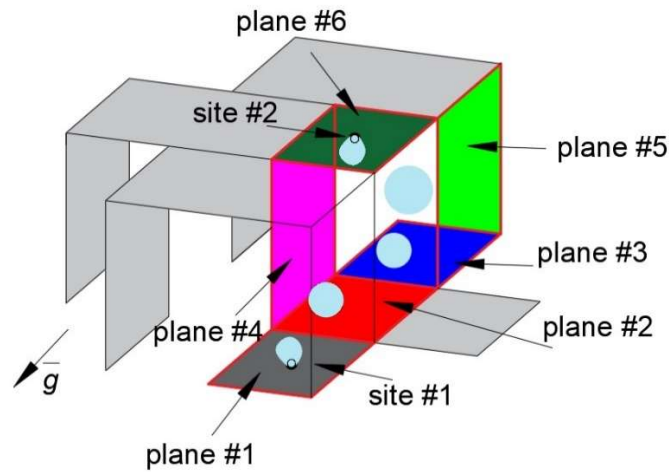

Fig. 13. Planes used for local heat transfer evaluation

The same procedure was used to calculate local heat transfer coefficients as was discussed previously in section 4.7. The results are shown in Fig. 14. For all three plots presented, points at $t$ = 0 s correspond to single-phase heat transfer (these are solutions at the last time step of single-phase simulations which are recycled as initial conditions for the two-phase flow runs).

If one considers heat transfer evolution for bulk superheat of 0 degrees, it may be seen that for plane #1 the value does not change significantly. This implies that conditions for bubble growth and departure stay consistent over time. The plots from Fig. 10 for nucleation site #1 also prove this point. The same conclusion could be made for heat transfer coefficient behavior on plane #1 for bulk superheat 2 degrees. For bulk superheat 4 degrees there is a decline in heat transfer after ~10 ms. This happens due to coalescence between the departed bubble and the newly nucleated one. Because of that, the merged bubble stays longer near the wall consuming more energy from the surrounding liquid. For next bubble (point at ~16 ms), no coalescence event is observed. At ~32 ms an increase in heat transfer for plane #1 attributes to the fact that the cavity is full of liquid (no vapor phase is left at this time). The situation is different for plane #6: increases in heat transfer coefficient (for bulk superheats 0 and 2 degrees after ~60 and 30 ms correspondingly) indicate that small-sizes bubbles are nucleated (could be seen in Fig. 10 as well). Thus, more liquid occupies the cavity causing the rise of heat transfer. For bulk superheat 4 degrees there is a coalescence event again (at ~10 ms) happening between the departed and newly nucleated bubbles which attributes to a decrease in heat transfer coefficient.

For plane #2 a decrease in heat transfer coefficient is observed for all three cases (starting from ~8 ms). This is the result of the numerical assumption of Boiling Algorithm – vapor phase is considered to have saturated temperature. During single-phase simulations bubbles initially nucleated in the domain work as heat sinks. As the code runs, this saturated temperature affects thermal boundary layer (areas of lower temperature near walls could be seen at 8 ms in Fig. 11 and Fig. 12). Although this effect distorts heat transfer evaluation at the beginning, it disappears as boiling simulations are run longer. For plane #3 one could clearly see the effect of bulk superheat on heat transfer: before bubbles reach this plane, the average heat transfer coefficient values are 114.7 (0 degrees), 189.5 (2 degrees) and 253.5 (4 degrees) W/(m$^2$·K).

For planes #4 and 5 the situation is similar: values are nearly constant until bubbles come closer to these planes causing disturbances in thermal and hydrodynamic boundary layers (e.g., see the growth on plane #4 after ~60 ms for bulk superheat 0 degrees, Fig. 14).

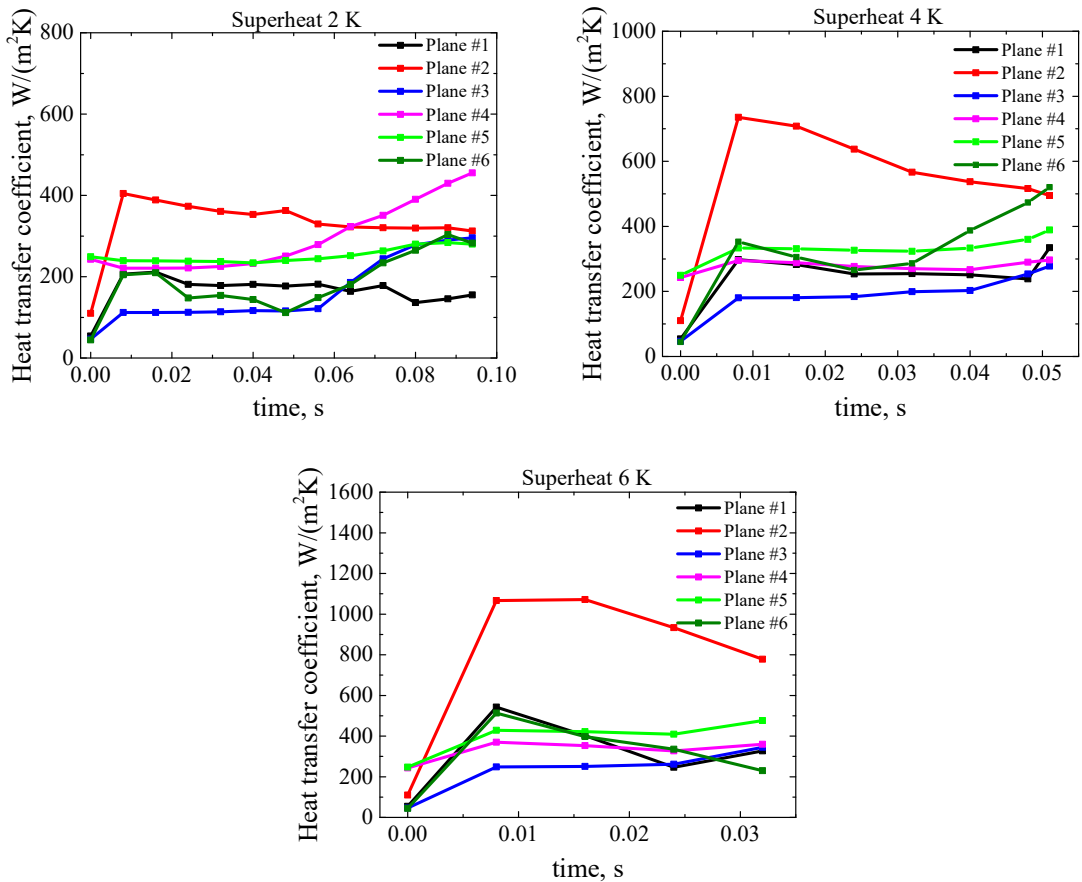

Fig. 14. Local heat transfer coefficient

## 6. CONCLUSIONS

Interface capturing flow boiling simulations are conducted for a vertical minichannel with offset strip fins (OSF) using code PHASTA. First, to establish the credibility of the results, model validation is conducted for a single nucleation site in a vertical rectangular channel. The observed bubble departure diameter agrees well with the experiments for the smallest cavity size considered. The steam generation and bubble departure frequency are underpredicted which could be attributed to microlayer evaporation (not considered in the current study).

Second, in a complex geometry with offset strip fins, single-bubble flow boiling was analyzed. Bubble dynamics characteristics are investigated with varying contact angle and wall superheat. It was observed that temperature distribution in the channel plays a major role in determining heat transfer process for this laminar flow ($Re$ = 200). The increase in bulk superheat by 2 degrees leads to a nearly 50% growth of the bubble departure diameter and the increase by 4 degrees allows to achieve almost a 170% growth.

Third, to observe multiple consequent bubble nucleation, two nucleation cavities were introduced in the same channel with OSF. Single-phase solutions are used as initial conditions for boiling simulations. Bulk superheat sensitivity tests are conducted to analyze bubble dynamics. Analysis of departure diameter values shows that the 4 degrees temperature difference in the bulk fluid may result in 40% difference in bubble departure diameter value. Local heat transfer coefficient is evaluated on each plane along the bubbles' path. It is shown that for some cases (e.g., for bulk superheat 0 degrees, plane #3) bubbles approaching the wall may increase heat transfer coefficient in 2.5 times. With a smaller nucleation diameter, more liquid is cooling the cavity which, in turn, results in nearly twice higher heat transfer coefficient (e.g., for bulk superheat 2 degrees, plane #6).

Future work will be focused on improving model predictions of bubble departure frequency and amount of steam released from a cavity as well as testing PHASTA capabilities on modeling flows with a higher void fraction.

**ACKNOWLEDGMENT**

Acusim linear algebra solution library is employed provided by Altair Engineering Inc. along with mesh and geometry modeling libraries provided by Simmetrix Inc for simulations. We also acknowledge the computing resources provided on Henry2, a high-performance computing cluster operated by North Carolina State University. Authors greatly appreciate the constructive and useful comments from the reviewer.

**FUNDING**

Current work is performed under Two-Phase Flow DNS Phase 2 Project supported by Mitsubishi Heavy Industries, Ltd.

**NOMENCLATURE**

$u_i$            fluid velocity, m/s

$\rho$            density, kg/m³

$p$            static pressure, Pa

$\tau_{ij}$            viscous stress tensor, Pa

$f_i$            body force (including gravity and surface tension), Pa/m

$c_p$            specific heat at the constant pressure, J/(kg·K)

$T$            temperature, K

$k$            thermal conductivity of a fluid, W/(m·K)

$\vec{q}$            dissipation function, J/(m³·s)

$t$            time, s

$H_\varepsilon$            smoothed Heaviside function

$\varepsilon$            interface thickness, m

$\rho_l$            liquid density, kg/m³

$\rho_g$            vapor density, kg/m³

$F_{CA}$            contact angle force, H

$F_1$            contact angle algorithm parameter

$T_{CA}$            thickness of contact angle force application region, m

$H_{CA}$    height of contact angle force application region, m

$d_{wall}$    distance from the wall, m

$R_0$    bubble radius, m

$R_1$    radius of temperature gradient collection outer shell, m

$\bar{q}$    volume-averaged heat flux, W/m$^4$

$h_{fg}$    latent heat of evaporation, J/kg

$V$    bubble volume, m$^3$

$t_b$    nucleation period, s

**SUPPLEMENTAL MATERIAL**

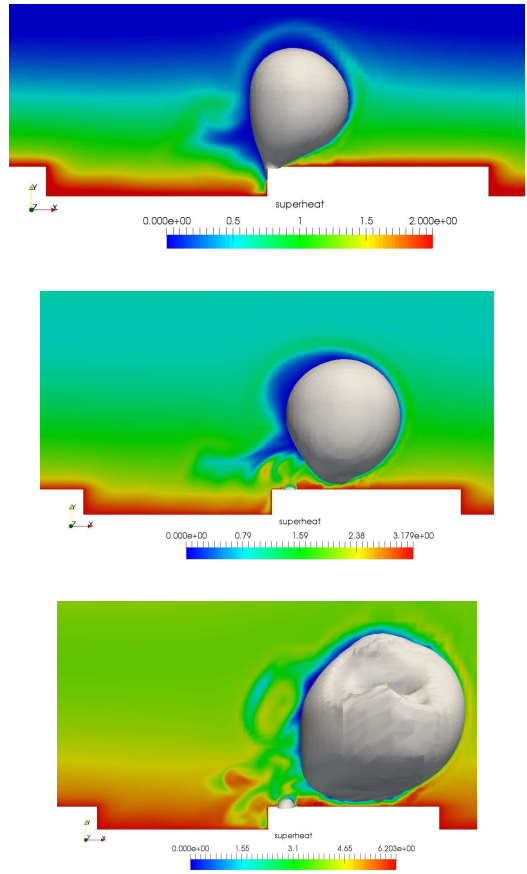

Figure S1. Bubbles location at 27 ms

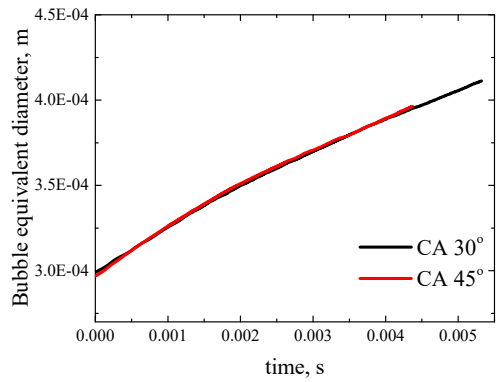

Figure S2. Bubble equivalent diameter growth for contact angles 30° and 45°

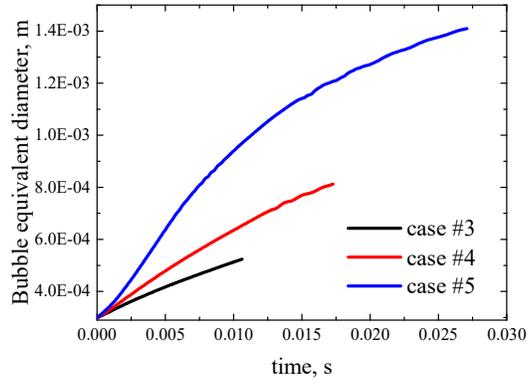

Figure S3. Bubble equivalent diameter growth for wall superheat 2, 4 and 6 K until the

departure moment

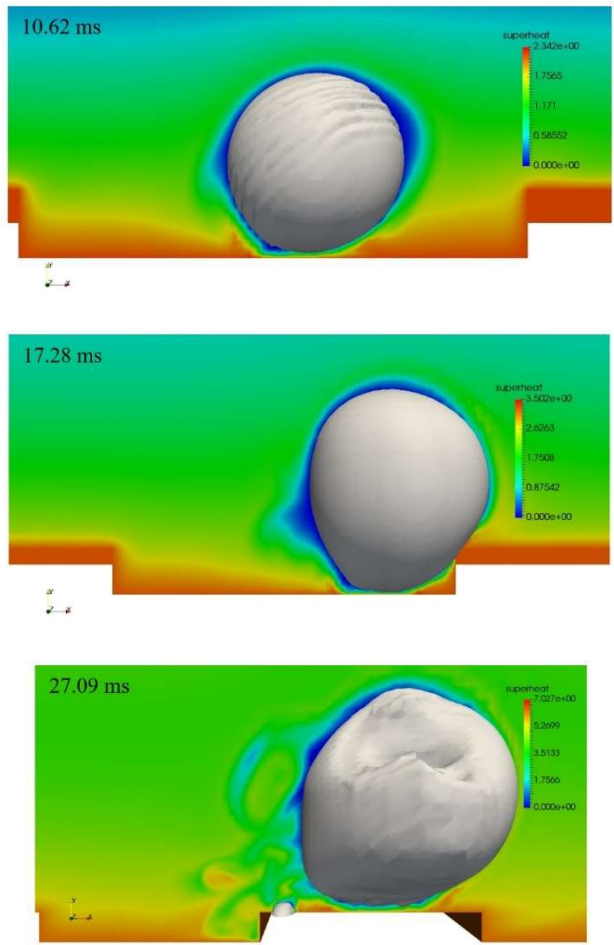

Figure S4. Bubble at the departure for cases #3-5 (from top to bottom)

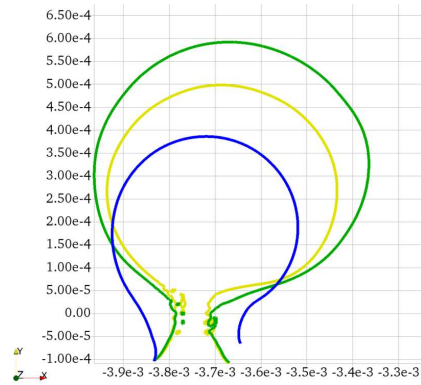

Figure S5. Bubble shapes at the right wall cavity for 2 (blue), 4 (yellow) and 6 (green) K

(scale in m)

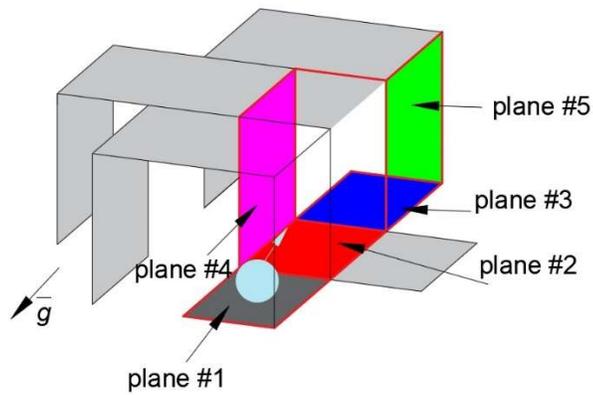

Figure S6. Plane locations used for the local heat transfer evaluation

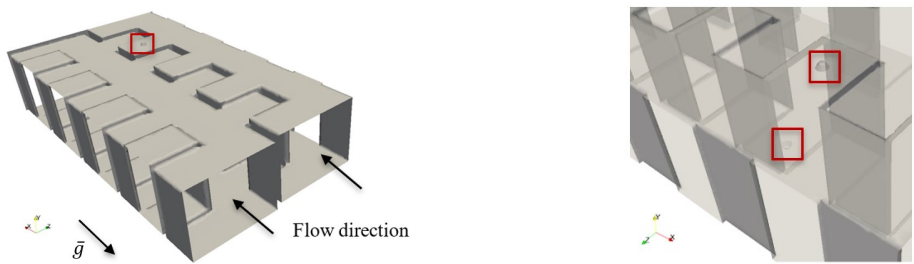

Figure S7. Simulation domain (cavity on the left channel wall is shown in red) (left) and

magnified view of cavities (right)

To properly account for boiling and, at the same time, maintain reasonable computational cost, several mesh refinement regions are created for bubble departure from nucleation cavities cases (Figure S8). Additionally, BLs are set on all channel walls to resolve friction losses. Mesh parameters are listed in Figure S9. Initial bubble shape (shown in yellow at the left wall cavity)

Table S1. Total number of finite elements is 6,212,594.

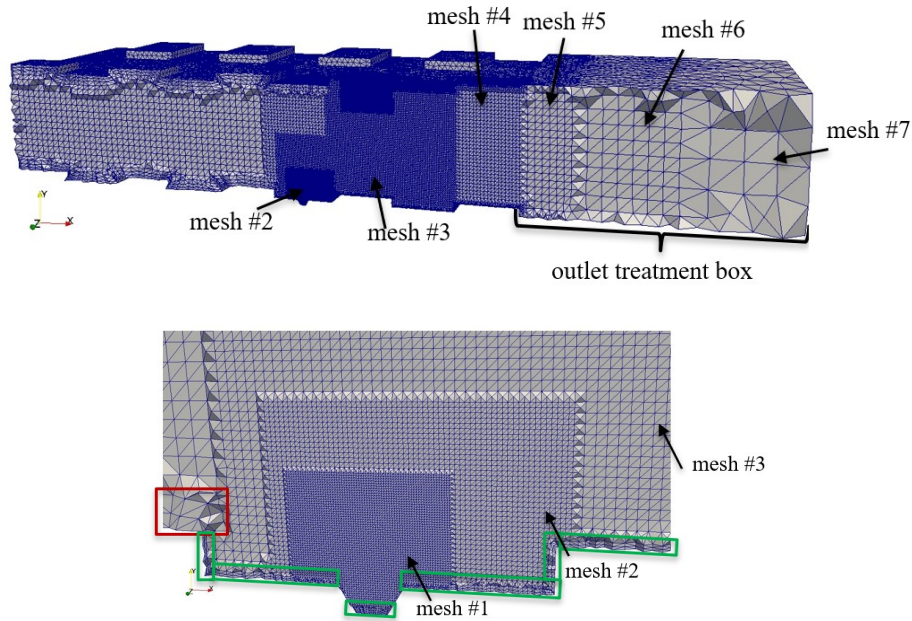

Figure S8. Mesh refinement regions (top), magnified view of the cavity at the right wall

(in red and green boxes different BLs are shown) (bottom)

Initial bubble diameter is 0.42 mm, it is created to have CA 90° (see Figure S9, the same bubble is initialized at the right wall cavity). This initial diameter is chosen to accurately resolve bubble growth (roughly 25 elements across the bubble diameter). Target CA is set as a range 10-20°.

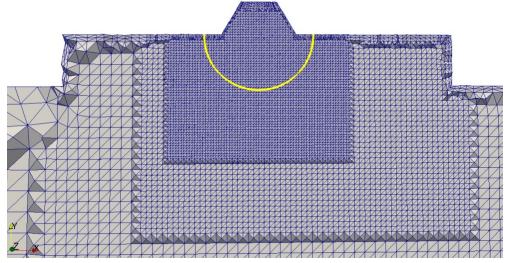

Figure S9. Initial bubble shape (shown in yellow at the left wall cavity)

Table S1. Parameters of mesh refinement regions

| # | Mesh resolution, m |
|---|---|
| 1 | $1 \cdot 10^{-5}$ |
| 2 | $2 \cdot 10^{-5}$ |
| 3 | $4 \cdot 10^{-5}$ |
| 4 | $8 \cdot 10^{-5}$ |
| 5 | $1.6 \cdot 10^{-4}$ |
| 6 | $3.2 \cdot 10^{-4}$ |
| 7 | $6.4 \cdot 10^{-4}$ |

**Bubble Trajectories**

Bubble growth (bubble size at each location is indicated with the color) with respect to their $(x, y, z)$ coordinates is presented in Figure S10. Bubbles dissipate in the coarse grid region near the outlet. However, some trajectories stop in the middle in the domain (e.g., see 3D view for superheat 6 K) because of the simulation time limitation. One could see that for all three cases as bubbles departed from cavities, they tend to bounce back to the walls (e.g., see 3D view for 2 and 6 K, Figure S10) and this effect is more pronounced for the right wall cavity. This bouncing motion has been previously observed in vertical channels [19, 20]. From top views one could conclude that small bubbles are more likely to deviated from the centerline whereas large-size bubbles tend to have a straighter trajectory (see top views for 2 K and 6 K superheats, Figure S10).

| 3D view | Top view |
|:---:|:---:|
| 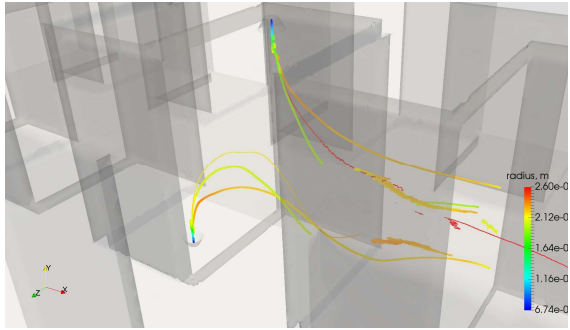 | 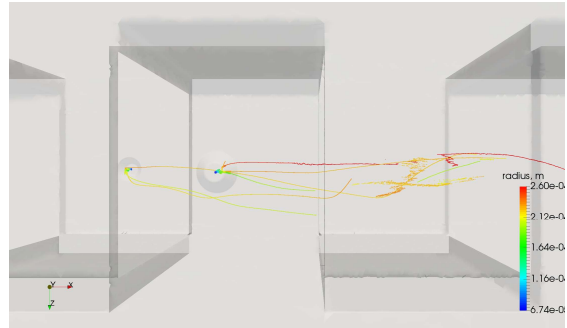 |

Superheat 2 K

| 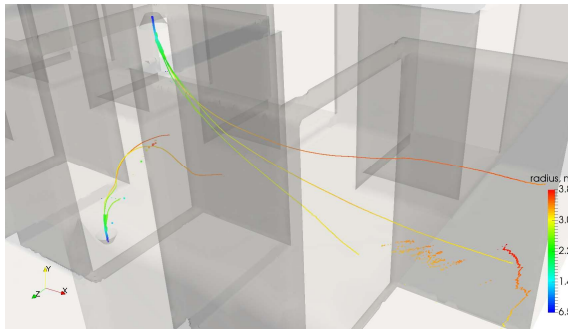 | 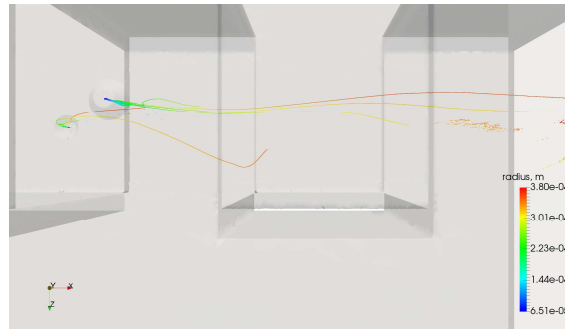 |
|:---:|:---:|

Superheat 4 K

| 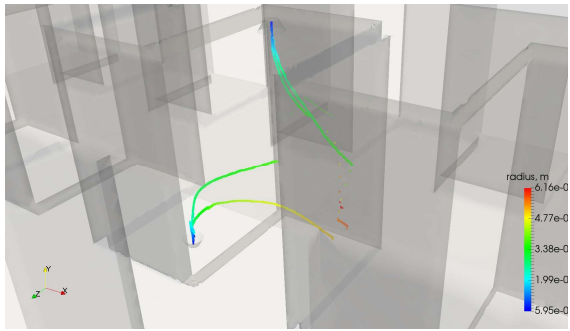 | 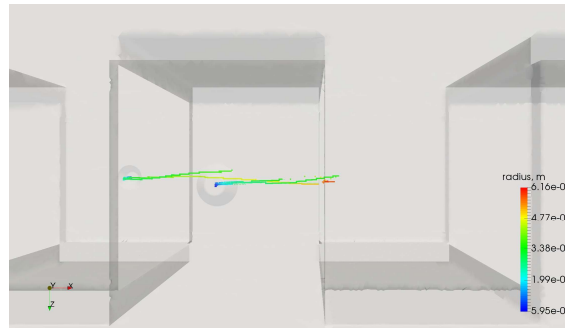 |
|:---:|:---:|

Superheat 6 K

Figure S10. Bubble pathways for different superheats

**Figure Captions List**

Fig. 1. Computational domain for validation studies

Fig. 2. Different cavity sizes considered: baseline (top), 38% smaller (middle) and 61% smaller (bottom)

Fig. 3. Bubble statistics are compared with experiments [20] (dash lines on the left plot define uncertainty in departure diameter measurements)

Fig. 4. Geometry of OSF (left), corresponding computational domain (right, dimensions are shown in m)

Fig. 5. Mesh refinement regions (top), magnified view of the mesh near the initialized bubble (bottom)

Fig. 6. Velocity (blue line) and temperature (green, black and red lines) profiles applied for three superheat cases

Fig. 7. Bubbles at initial time step (top) and at the departure moments (bottom)

Fig. 8. Local void fraction evaluation region (in red) (left) and void fraction statistics for cases #3-5 (right)

Fig. 9. Local heat transfer coefficient changes for different planes on the bubble path

Fig. 10. Bubble statistics for nucleation site #1 (on the right wall, open symbols) and #2 (on the left wall, solid symbols) for 0 (red), 2 (black) and 4 (blue) degrees bulk superheat

Fig. 11. Bubble dynamics for 4 degrees bulk superheat

Fig. 12. Bubble dynamics for 0 and 2 degrees bulk superheat

Fig. 13. Planes used for local heat transfer evaluation

Fig. 14. Local heat transfer coefficient

**Table Caption List**